\documentclass[AMA,STIX1COL]{WileyNJD-v2}
\usepackage{moreverb}
\usepackage{cancel}
\usepackage{color}
\newcommand\BibTeX{{\rmfamily B\kern-.05em \textsc{i\kern-.025em b}\kern-.08em
T\kern-.1667em\lower.7ex\hbox{E}\kern-.125emX}}

\articletype{RESEARCH ARTICLE - FUNDAMENTAL}%

\received{<day> <Month>, <year>}
\revised{<day> <Month>, <year>}
\accepted{<day> <Month>, <year>}

\begin{document}

\title{ A hybrid echocardiography-computational fluid dynamics framework for ventricular flow simulations}

\author[1]{Mohammadali Hedayat}

\author[2]{Tatsat R. Patel}

\author[3]{Marek Belohlavek}

\author[4]{Kenneth R. Hoffmann}

\author[1]{Iman Borazjani*}

\authormark{Hedayat \textsc{et al}}

\address[1]{\orgdiv{J. Mike Walker '66 Department of Mechanical Engineering}, \orgname{Texas A\&M University}, \orgaddress{\city{College Station}, \state{TX}, \country{USA}}}

\address[2]{\orgdiv{Department of Mechanical and Aerospace Engineering}, \orgname{State University of New York at Buffalo}, \orgaddress{\city{Buffalo}, \state{NY}, \country{USA}}}
	
\address[3]{\orgdiv{Department of Cardiovascular Diseases}, \orgname{Mayo Clinic}, \orgaddress{\city{Scottsdale}, \state{AZ}, \country{USA}}}

\address[4]{\orgdiv{Department of Neurosurgery}, \orgname{University at Buffalo SUNY}, \orgaddress{\city{Buffalo}, \state{NY}, \country{USA}}}

\corres{*Iman Borazjani, J. Mike Walker '66 Department of Mechanical Engineering, Texas A\&M University. \email{iman@tamu.edu}}


\abstract[Abstract]{Image-based computational fluid dynamics~(CFD) has emerged as a powerful tool to study cardiovascular flows while 2D echocardiography~(echo) is the most widely used non-invasive imaging modality for diagnosis of heart disease. Here, echo is combined with CFD, i.e., an echo-CFD framework, to study ventricular flows. To achieve this, our previous 3D reconstruction from multiple 2D-echo at standard cross-sections is extended by 1) reconstructing valves~(aortic and mitral valves) from 2D-echo images and the superior wall; 2) optimizing the location and orientation of the surfaces to incorporate the physiological assumption of fixed apex as a reference~(fixed) point for reconstruction; and 3) incorporating several smoothing algorithms to remove the nonphysical oscillations~(ringing) near the basal section observed in our previous study. The main parameters of the reconstructed left ventricle~(LV) are all within \textbf{}the physiologic range. Our results show that the mitral valve can significantly change the flow pattern inside the LV during the diastole phase in terms of ring vortex formation/propagation as well as the flow circulation. In addition, the abnormal shape of unhealthy LV can drastically change the flow field during the diastole phase. Furthermore, the hemodynamic energy loss, as an indicator of the LV pumping performance, for different test cases is calculated which shows a larger energy loss for unhealthy LV compared to the healthy one.}


\maketitle


\section{Introduction}\label{sec1}
In the U.S. more than $375,000$ deaths are associated with heart diseases each year \cite{american2015heart}. Early diagnosis of heart disease can reduce the mortality rate and provide useful information on proper therapies. Several noninvasive techniques have been used to characterize abnormal blood patterns that contribute to heart failure. Medical imaging techniques including Phase Contrast-MRI (PC-MRI)~\cite{bock20104d,wigstrom1999particle}, 3D echocardiography~(3D-echo)~\cite{sugeng2007real,chaoui2004three}, and cardiovascular magnetic resonance~(CMR)~\cite{rodriguez2013intracardiac} have been widely used to visualize flow patterns in the cardiac system. However, they typically suffer from low spatial/temporal resolution within an acceptable scanning time and thus insufficient for detailed hemodynamic analysis~\cite{doost2016heart}. Computational Fluid Dynamic~(CFD) in combination with imaging techniques for geometric reconstruction can provide a powerful tool for investigating the flow pattern with more details in the reconstructed cardiac system~\cite{borazjani2013left}.


The flow simulations in the complex geometries with large deformations and complicated motions such as left ventricle~(LV) depends mainly on the geometry and its motion. The geometric segmentation and reconstruction process from medical images along with the assumptions for motions of these geometries~(valves and LV) can greatly influence LV's flow pattern~\cite{mittal2016computational,votta2013toward,le2013fluid}. Currently, two types of techniques are available to model the motion of valves and the LV~\cite{doost2016heart}: 1) Fluid-Structure Interaction~(FSI), and 2) prescribed models. In prescribed models, the motions and geometries are prescribed
based on in vivo measurements ~\cite{saber2003progress,long2008subject,schenkel2009mri} or simplified equations to define the motion~\cite{domenichini2005three,le2013fluid,seo2013effect,azar2018mechanical}. Of course, obtaining the motions from medical images is more realistic than prescribing the motion based on simplified equations. The FSI models have been widely used in numerical simulations~\cite{watanabe2004multiphysics,cheng2005fluid,borazjani2008curvilinear,borazjani2013fluid,sharzehee2018fluid,kamensky2015immersogeometric,shahidian2017stress} especially for capturing the motion of heart valves. However, due to the complexity of geometry, dynamic shape, and large deformation of LV and mitral valve~(MV) leaflets FSI models are computationally expensive~\cite{borazjani2015review,mittal2016computational}, and the lack of data for tissue mechanical properties from in vivo measurements makes using them quite challenging. In fact, FSI models are typically validated by comparing the motion computed from FSI with the in vivo measurements, i.e., FSI methods are considered valid when their motion is similar to prescribed ones. Therefore, it is desirable to use image-based geometry/motion if available. However, the main concern about using the medical images in CFD simulations is their accuracy in terms of spatio-temporal resolution.

 Several imaging techniques are currently available for LV, e.g., computed tomography~(CT)~\cite{watabe2015enhancement,szafer2019simplified}, magnetic resonance imaging~(MRI)~\cite{lapa2015imaging}, cardiac magnetic resonance~(CMR)~\cite{morris2018magnetic}, and 3D-echo~\cite{monaghan2006role}. However, most of these techniques are either very expensive or suffer from the lack of spatial or temporal resolution within an acceptable acquisition time. Among these, 2D-echo is the most widely used non-invasive imaging technique, due to its low costs and fast acquisition ~\cite{chamsi2017handheld}. 2D echo can provide high temporal resolution~(from 250 fps, down to 50 fps for resolving the whole LV) compared to other techniques~\cite{rajan2016automated}. However, the acquired data from segmentation of 2D echo are only available on a few standard planes from which the 3D geometry need to be reconstructed~\cite{rajan2016automated}. 
 
For CFD simulations of LV, the valves need to be modeled. Due to the highly dynamic motion and geometry of the leaflet valve, the reconstruction of heart valves still remains a challenge. In many studies, the valve's leaflets are either ignored or simplified using an on-off approach where the switch between the on and off configuration occurs instantaneously without any intermediate positions~\cite{song2015role,watanabe2008looped,zheng2012computational,domenichini2005three}. Dimasi et al.~\cite{dimasi2012influence} reconstructed the MV and AV from cardiac magnetic resonance~(CMR) images. Their study was limited to systole~(MV always closed). Seo eta al.~\cite{seo2014effect} created a prescribed kinematic model for MV motion based on the location of the tip of the valve leaflets from echo images for a cardiac cycle. However, mitral annulus was assumed to be a perfect circle with a fixed dimension in their study. Chnafa et al.~\cite{chnafa2014image} modeled the MV based on a projection of mitral annulus geometric with the assumption of elliptical MV orifice, and the aortic valve~(AV) was approximated by a simple plane. A reduced degree of freedom model for leaflet motion coupled the left ventricle geometry was introduced by Domenichini et al.~\cite{domenichini2015asymptotic}. However, the mitral annulus was assumed to be a rigid circular orifice in their study. Su et al.~\cite{su2016cardiac} reconstructed MV using a mathematical model based on only one cross section from CMR. Here, we reconstruct mitral and aortic valves using the segmented data from multiple axis 2D echo images with the temporal resolution of more than $30$~fps.

The valve geometry and leaflet motion can have a significant impact on the flow field inside the LV~\cite{seo2014effect}. The effect of AV has been studied thoroughly in terms of hemodynamic performance~\cite{seaman2014steady,borazjani2010high}, platelet activation~\cite{hedayat2019comparison,hedayat2017platelet,alemu2007flow} and thrombus formation~\cite{bluestein2000vortex,yoganathan2005flow}, both experimentally and numerically. Most of these simulations are performed in simplified geometries without an LV. Only a limited number of studies considered the effect of ventricular valves in their LV simulations~ \cite{mao2017fully,su2015numerical,le2013fluid,su2014numerical,su2016cardiac,chnafa2014image,seo2014effect,dahl2012fsi,carmody2006approach}. Due to the significant impact of MV on the performance of the heart, most of LV simulations are just considering the effect of mitral valves~(neglecting the AV) and the simulations are mainly limited to the diastolic phase  \cite{su2015numerical,su2016cardiac,chnafa2014image,seo2014effect,dahl2012fsi,carmody2006approach}. Charonko et al.~\cite{charonko2013vortices} reported that the mitral vortex ring facilitates the filling and enhances flow transfer to the LV apex. Dahl et al.~\cite{dahl2012fsi} studied the effect of MV in 2D simulation of LV. Their results show that asymmetric leaflets for MV as well as an adequate model for the left atrium are essential for resolving important flow features in the LV. Seo et al.~\cite{seo2014effect} showed the MV to significantly affect the vortex ring propagation and flow field inside LV. They also found that due to the asymmetry of the MV a circulatory flow pattern can be generated in the LV which can enhance apical washout ability and reduce the risk of thrombus formation. Su et al.~\cite{su2016cardiac} investigated the effect of MV on vortex formation time~(VFT). Their results showed that VFT is a promising parameter to characterize the performance of the LV. However, only a few simulations studied the flow pattern inside the LV incorporated with both ventricular valves~\cite{su2014numerical,su2016cardiac,mao2017fully} and even fewer compared flow pattern in healthy LV with left ventricles with any kind of heart failure~\cite{su2016cardiac}. 

In this study, a generalized method to automate reconstruction of the 3D LV geometry from standard 2D echo projections~\cite{rajan2016automated} is extended to reconstruct the valves and the superior wall to generate a closed 3D geometry for CFD simulations~(section~\ref{Superior wall reconstruction}). A new method is developed for the reconstruction of AV and MV based on $1$ and $3$ standard sections from 2D echo, respectively (section~\ref{Valves reconstruction}). This 3D reconstruction method is coupled with our in-house CFD code based on a sharp-interface immersed boundary method \cite{gilmanov2005hybrid,borazjani2008curvilinear,borazjani2013parallel} to simulate the ventricular flows (section~\ref{Hybrid echo-CFD framework}). Different smoothing algorithms, described in section~\ref{Data smoothing}, are tested to dampen the fluctuations of boundaries obtained from echo data. The effect of these smoothing algorithm on shape and flux of LV as well as CFD results are investigated~(section \ref{smoothness in geometric} and \ref{numerical simulations smoothing}). This hybrid echo-CFD framework is applied to a healthy LV and an LV with acute myocardial infarction~(AMI) and the results are discussed in section~\ref{Comparing healthy and AMI}.

\section{Methods}
In this section, the previous process for the reconstruction of LV from multiple 2D echo cross-section~\cite{rajan2016automated} is briefly described in \ref{Previous LV reconstruction method} subsection. Afterwards, the new developments/modifications to this method are discussed in \ref{Data smoothing} to \ref{Computational setup} subsections. The new 3D reconstruction method is applied to a healthy LV as well as one with acute myocardial infarction~(AMI) which was scanned after inducing AMI on the healthy LV by ligating the mid-portion of the left anterior descending coronary artery~(both healthy and AMI images are approved by the Institutional Animal Care and Use Committee at the Mayo Clinic).

\subsection{Previous LV reconstruction method} \label{Previous LV reconstruction method}
The following steps are performed by Rajan et al.~\cite{rajan2016automated} to reconstruct the LV from 2D-echo images: 

\begin{enumerate}[I]
	\item \textbf{Endocardial border detection}\\
	LV endocardial borders are identified using the gradient in the RGB~(Red, Green, Blue) values and extracted from black and white echo image in standard long axis~(LA) and short axis~(SA) echo projections of a pig heart for an entire cardiac cycle with the frame rate ranged from $36$ fps to $55$ fps for different sections.
	
	\item \textbf{B-spline data smoothing}\\
	The segmented boundaries extracted from echo images are not smooth and a B-spline curve fitting is used for smoothing the detected endocardial borders.
	
	\item \textbf{Temporal interpolation}\\
	Since the image frame rate for various sections is not the same, temporal interpolation is used to obtain different cross-sections at the same time instants. A cubic spline with natural boundary conditions is used for temporal interpolation~\cite{rajan2016automated}.
	
	\item \textbf{Scaling, positioning and orientation of the LV sections}\\
	The left ventricular end-diastolic dimension (LVEDD) is chosen as the scaling parameter to scale all the extracted data to the physical dimension. The angular orientations and position of the LA and SA sections are optimized using an optimization algorithm based on the assumption of a fixed apex.
	
	\item \textbf{Spatial Interpolation}\\
 	The available optimized cross-sections are interpolated using a bivariate method (depending on the available sections) to generate surface points at the chosen spatial resolution.
	
	\item \textbf{Temporal Smoothing}\\
	To make the geometry suitable for CFD applications, the volume flux is smoothed in time using the Fourier curve fitting method considering the periodic nature of the dependent variable.
	
	\item \textbf{Mesh Generation and Output}\\
	The final surface points are turned into a triangular mesh as needed by the immersed boundary~(IB) node classification algorithm in our in-house the CFD solver\cite{borazjani2008curvilinear}.
\end{enumerate}
The following optimizations are performed on the previous reconstruction method~\cite{rajan2016automated} to improve the accuracy and as well as incorporating physiological assumptions:

\subsection{Data smoothing} \label{Data smoothing}
\begin{figure}
	\centering
	\includegraphics[width=0.5\textwidth]{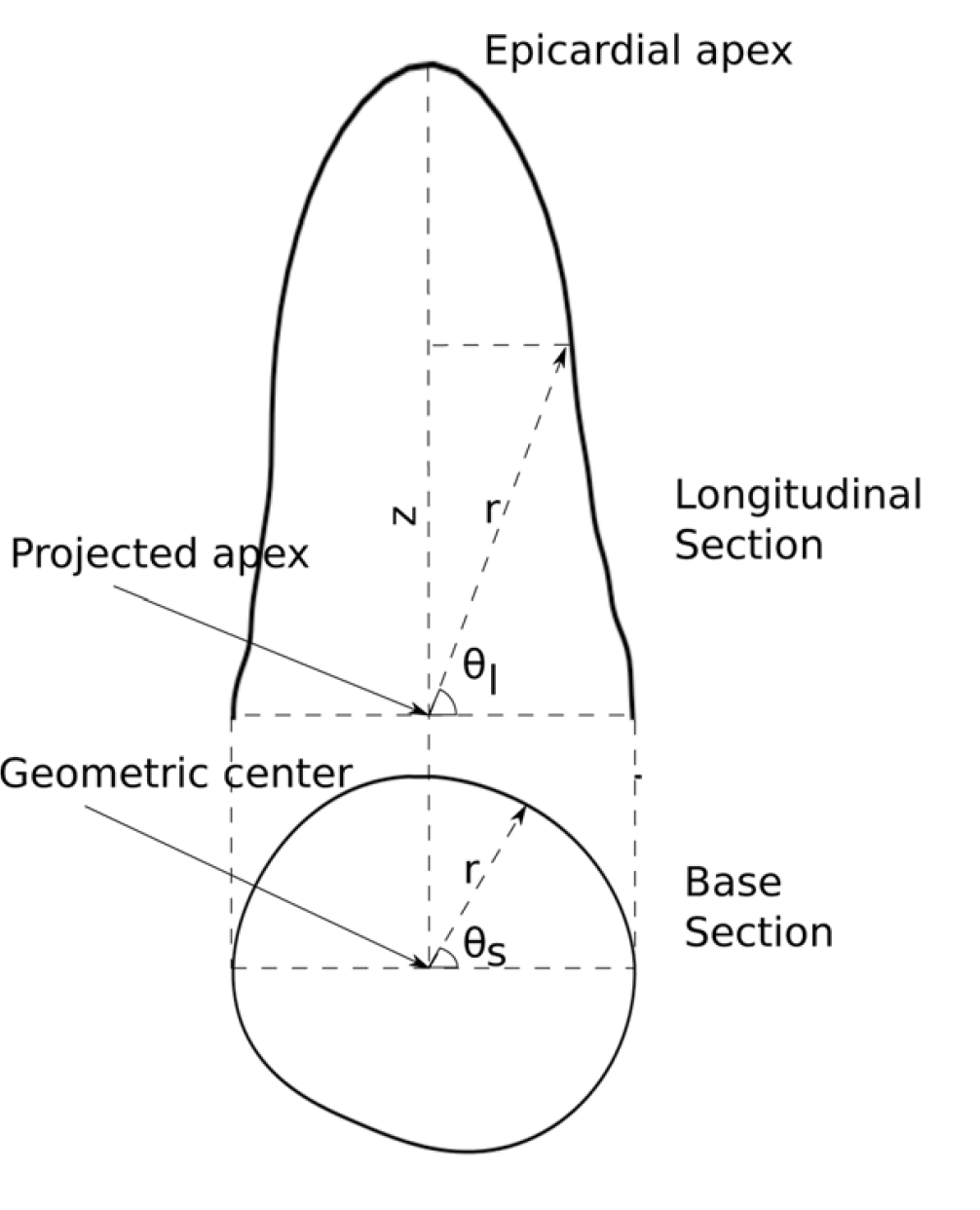}
	\caption{}
	\label{r-theta}
\end{figure}

\begin{figure} 
	\centering
	\includegraphics[width=1.0\textwidth]{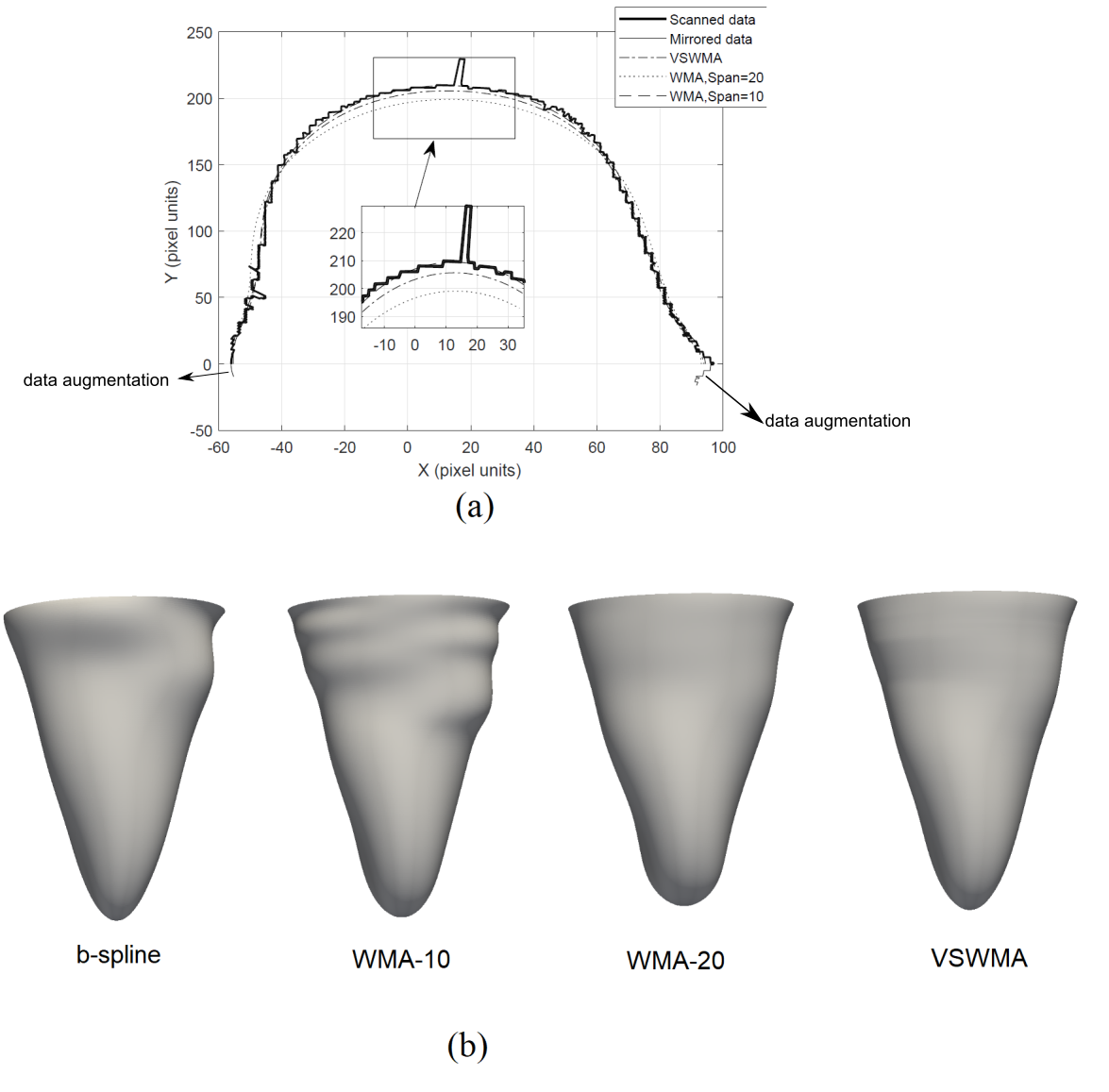}
	\caption{Comparison of effect of various smoothing techniques at early systole a) on the scanned data from three-chambered view Apical LA section at early systole b) final LV reconstruction}
	\label{smoothing}
\end{figure}

Rajan et al.~\cite{rajan2016automated} used a b-spline curve fitting method which resulted in small non-physical ringing near the basal section in final reconstructed LV's surface. To get rid of this ringing in the reconstructed geometry, the weighted moving average~(WMA), and finally, the variable span weighted moving average~(VSWMA) have been employed. Using WMA method, data at each node is computed as the weighted mean of the data points on both sides as
\begin{equation}
r(i)=\sum_{j=i-k}^{j=i+k} \frac{(k-j) \times r(j)}{2k+1}
\end{equation}
where $\Delta \theta=2k + 1$ is the span of the moving average scheme, $i$ is each node of data and $r(i)$ is radial coordinate of each node in $(r,\theta)$ format~(Fig.~\ref{r-theta}). Two general techniques are available for boundary point treatment when using moving averages methods: 1) the span is reduced progressively while reaching the boundary points in a way that the equation holds for all the data points; and 2) data augmentation is performed to generate a new dataset equivalent to the size of the span at the endpoints~(boundary points). Our results show that reducing the span cannot eliminate oscillations near the endpoints of reconstructed LV. Thus data augmentation is employed by reflecting data~(the number of data points to be mirrored is equal to the span size) around the boundary point~(Fig.~\ref{smoothing}-a). Different span sizes are tested and the effects of each can be seen in Fig.~\ref{smoothing}. Furthermore, Variable Span Weighted Moving Average (VSWMA) is employed, which enables us to apply larger span value near the endpoints and a smaller span near the apex. The span of the moving average scheme can vary based on the radial proximity of the data point to the vertical projection of the apex on the base~(as seen in LA apical sections Fig.~\ref{r-theta}) as follows: 
\begin{equation}
k(i)=k_s+(k_l-k_s)\times \frac{r_l-r(i)}{r_l-r_s}
\label{span}
\end{equation}
where $k(i)$, $k_l$, $k_s$, $r_l$, $r_s$ and $r(i)$ are the span value on the $i^{th}$ data point, the larger span, the smaller span, the largest $r$, the smallest $r$ and the radial distance of the $i^{th}$ data point from the mid-point of the endpoints of the dataset, respectively. Equation~\ref{span} basically interpolates the value of span in a way to have a larger span near the base and a smaller span near the apex.

\subsection{Scaling, positioning and orientation of the LV sections} \label{Scaling}

\begin{figure} 
	\centering
	\includegraphics{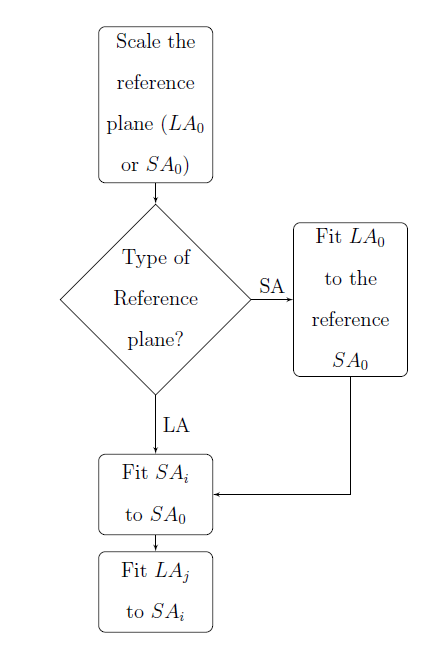}
	\caption{LV sections orientations, scaling and positioning algorithm}
	\label{optimization}
\end{figure}

\begin{figure} 
	\centering
	\includegraphics[width=1.0\textwidth]{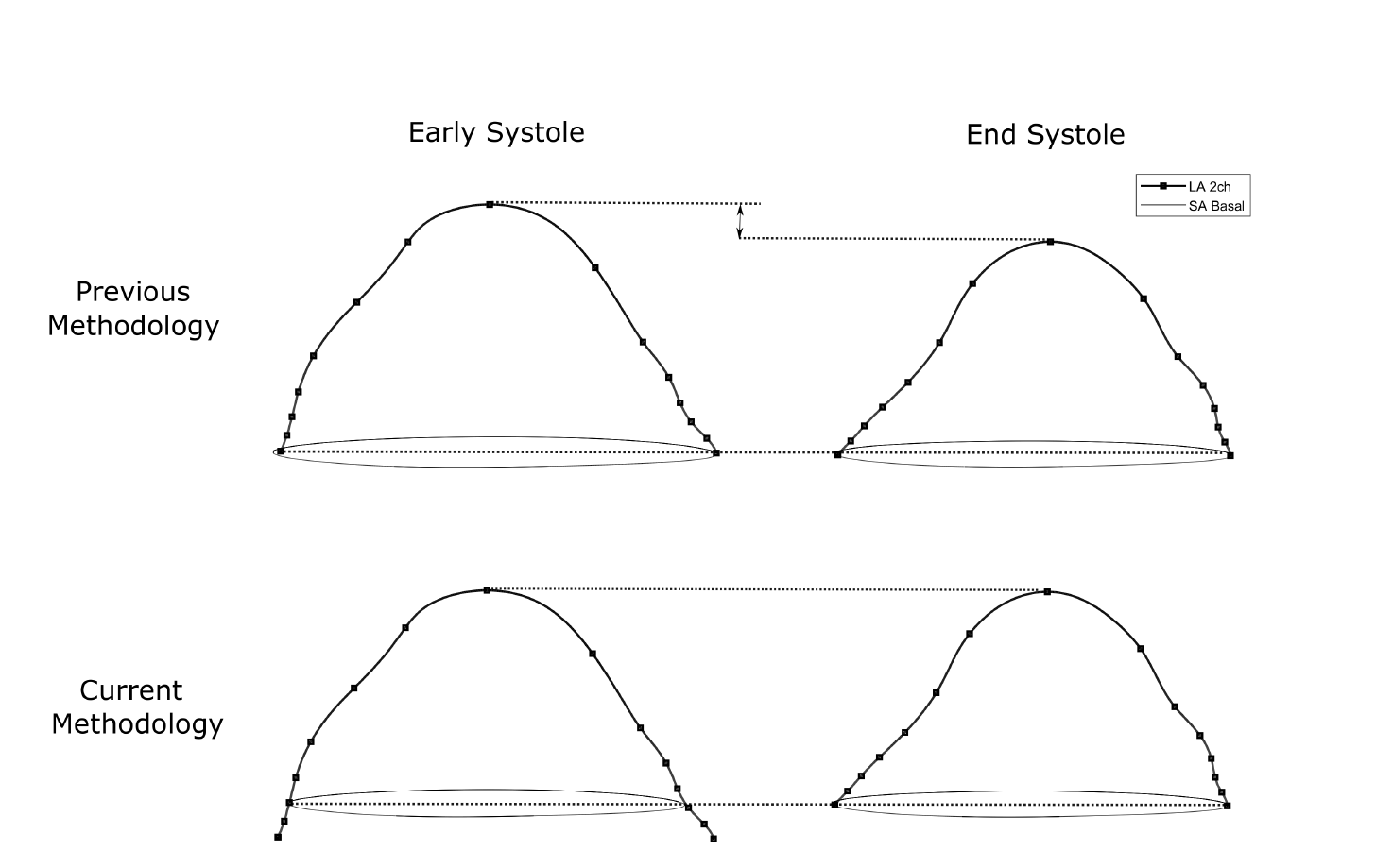}
	\caption{Comparison of the sectional scaling, positioning and orientation with the
		previous and current methodologies. Apical two-chambered LA section and SA Basal
		section are the sections in question.}
	\label{LV_orientation}
\end{figure}

The left ventricular end-diastolic dimension (LVEDD) is chosen as the scaling parameter to scale all the segmented data. The algorithm employed for fitting the SA and LA sections in our previous work~\cite{rajan2016automated} can be seen in Fig.~\ref{optimization}. The general algorithm has been kept the same with changes only to the positioning of the SA sections and the apex. Rajan et al.~\cite{rajan2016automated} assumed that the base is fixed while the apex moves in space and for the reconstruction, the LA sections were fit to the SA sections such that the ends of the LA Apical sections were fit to the SA basal sections. However, in-vivo experiments have shown that the longitudinal displacement of LV's apex is small compared to other parts of LV during a cardiac cycle~\cite{rogers1991quantification}. In addition, it is more realistic to assume that the SA sections are fixed in space for the 3D reconstruction because sections the images are captured by the transducer at fixed standard locations in the experiments. To that end, instead of varying the spatial position of the apex and the relative positions of the SA sections, the following is implemented and depicted illustratively in Fig.~\ref{LV_orientation}. The position of SA sections, as well as the apex, are assumed to be fixed during a cardiac cycle. The positions of the SA sections are calculated at end systole where the length of the left ventricle is minimum and are assumed to be fixed during a cardiac cycle. Therefore, the distances of the SA sections are relatively constant with respect to one another at all time instances.

\subsection{Valves reconstruction} \label{Valves reconstruction}
\begin{figure}
	\centering 
	\includegraphics[width=0.5\textwidth]{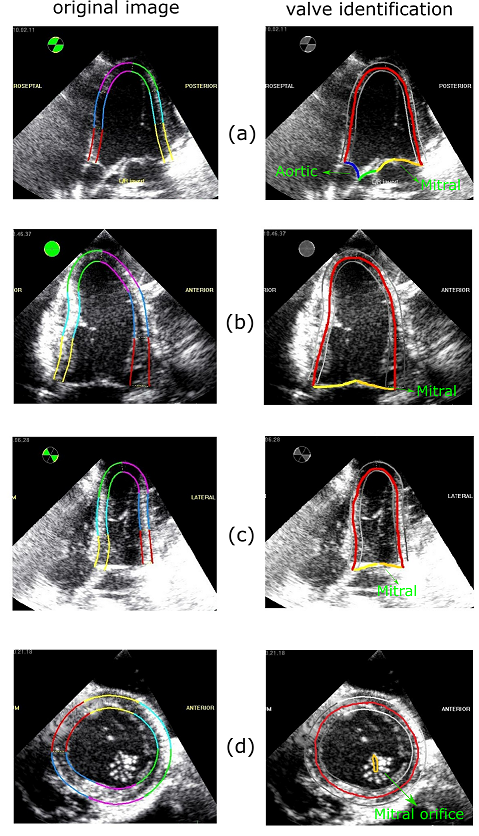}
	\caption{View of heart valves from different planes of 2D-echo at the beginning of the systole phase ($t=0$). (a) LA Apical three-chambered view (b) LA Apical two-chambered view (c) LA Apical four-chambered view (d) SA basal section.}
	\label{valve_ident}
\end{figure}

\begin{figure}
	\centering 
	\includegraphics[width=1.0\textwidth]{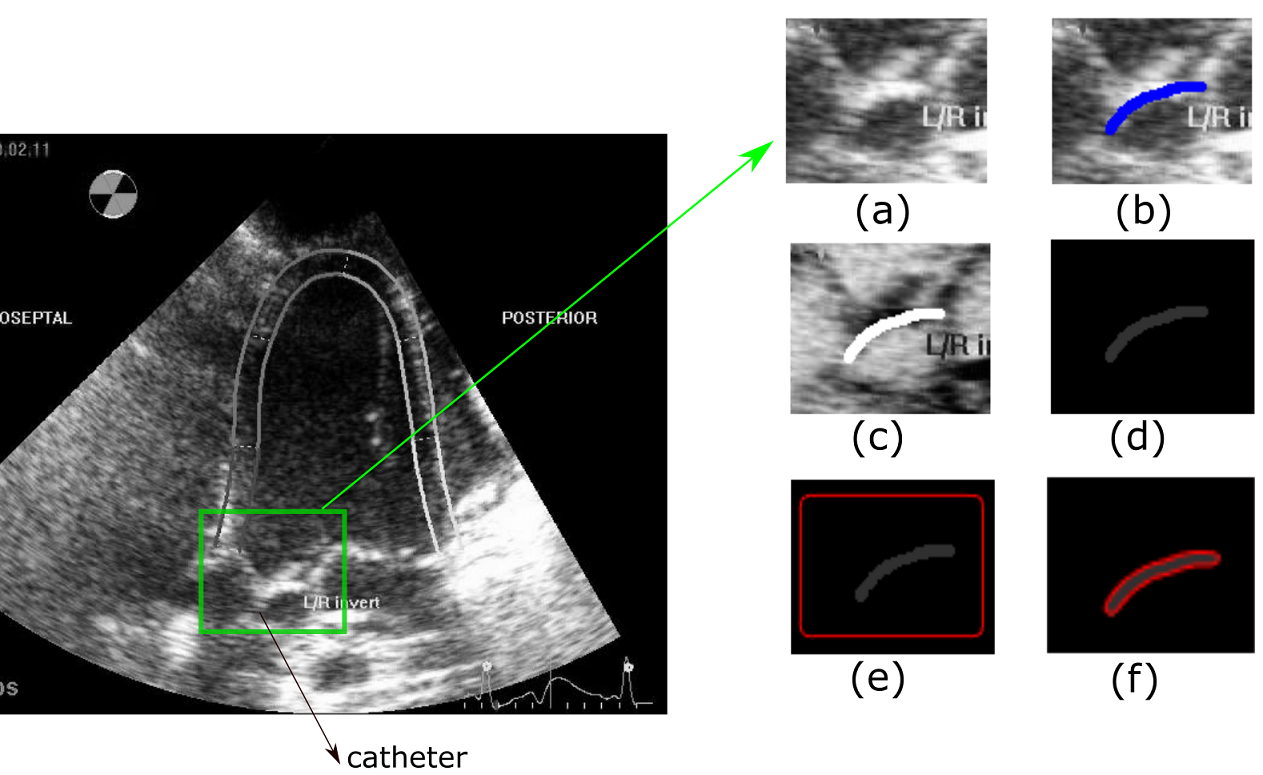}
	\caption{Steps taken for developing input videos for mitral and aortic valve reconstruction. All images are for t=0, LA Apical three-chambered view for one of the mitral valve leaflets. LA sections (a) Original image (grayscale) (b) Colored lea
		et using ImageJ (c) Step 1 of 2 in image manipulation (d) Step 2 of 2 in image manipulation (e) Initial LSF (f) Segmented image.}
	\label{LSF}
\end{figure}

\begin{figure} 
	\centering
	\includegraphics[width=0.8\textwidth]{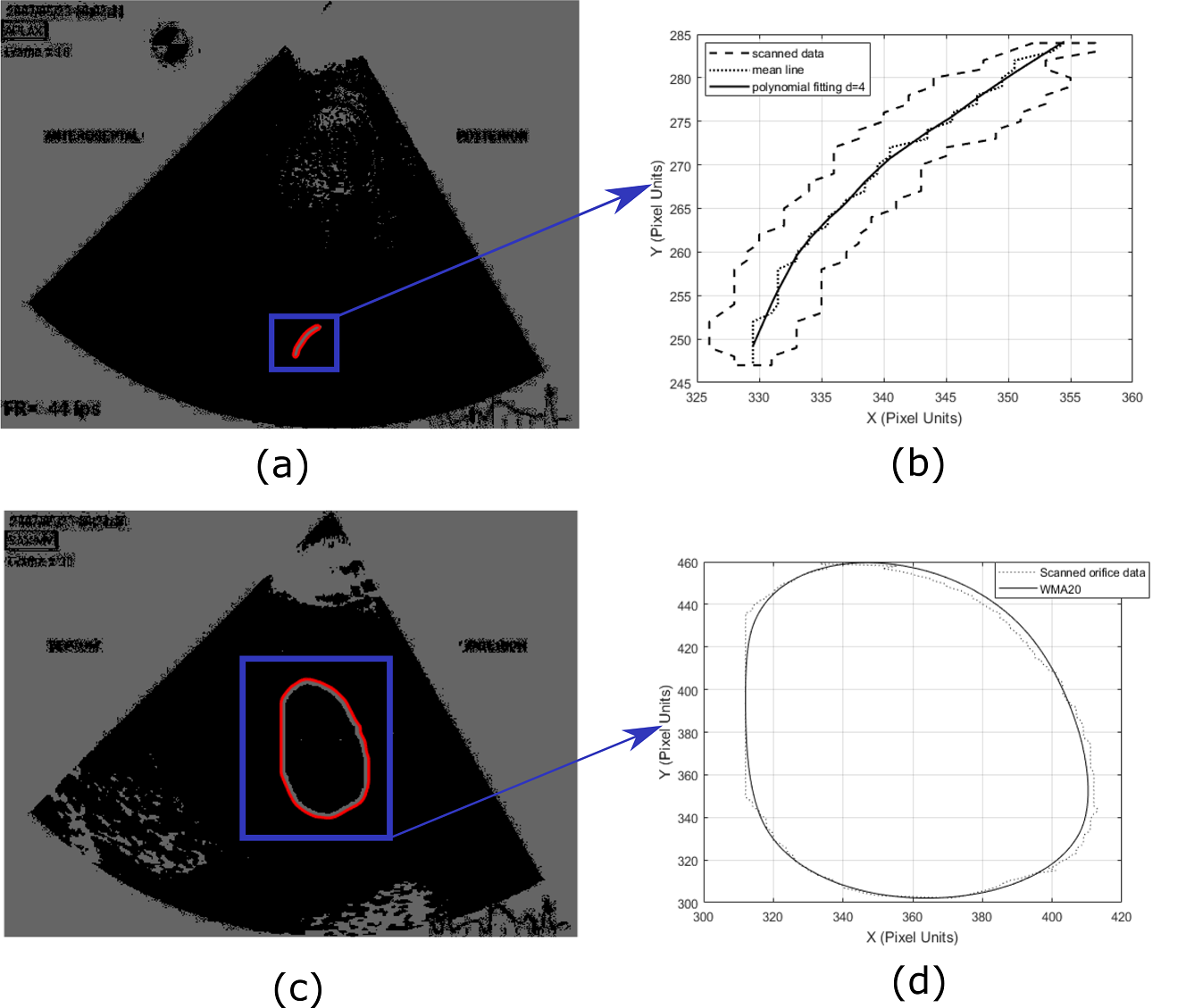}
	\caption{(a) Valve leaflet segmented image, (b) Leaflet scanned data, mean line and
		polynomial curve fitting d=4. (c) Valve orifice segmented image, (d) Orifice scanned data and WMA20 smoothed data.}
	\label{valve}
\end{figure}

In the LA Apical three-chambered view, both the mitral and aortic valves can be clearly seen~(Fig~\ref{valve_ident}-a). In the LA Apical two-chambered~(Fig~\ref{valve_ident}-b) and Apical four-chambered views~(Fig~\ref{valve_ident}-c), the mitral leaflets can be observed and the SA basal section the mitral orifice is clearly visible~(Fig~\ref{valve_ident}-d). The boundaries for valve leaflets and orifice on all the frames~(approximately at 30 fps) are manually identified using ImageJ software~\cite{schindelin2015imagej}~(Fig~\ref{valve_ident}). After the color pixelation, the following two manipulations are made to the images:

1) Inverting the 'Red' pixel values~($I_R$)~(Fig~\ref{LSF}-b),
\begin{equation}
I_R=255-I_R
\label{Inverting}
\end{equation}
considering that the images are in $8$-bit RGB format, and,

2) Intensity reduction~(Fig~\ref{LSF}-c)
\begin{equation}
I_R=
\left\lbrace 
\begin{tabular}{cc}
$0$ & if~ $I_R-250<0$\\
$40$ & otherwise 
\end{tabular}
\right.
\label{Intensity}
\end{equation}
Since the echo images are in grayscale all the pixels have approximately the same R, G, and B values. However, the colored pixels, the RGB intensities are not similar. The valves' leaflet are identified by the blue color, the $I_R$ from Step $1$ is close to $255$~($250$ for blue values is considered as the threshold, however, other values close to $255$ also can be used) and the above filter can assign every pixel valve in the image to $0$ except for the leaflet~(Fig~\ref{LSF}-d). It is to be noted that a similar computation shown in Eqn.~\ref{Inverting} and \ref{Intensity} can be done with 'G' instead of 'R' in the RGB pixel values as well to get a similar result. The next step in the input video development is to use the distance regularized level-set evolution~(DRLSE)~\cite{li2010distance} image segmentation for all the images of the same video~(Fig~\ref{LSF}-e) to segment the valve's leaflets in each frame. The initial level-set contour has also been depicted in Fig.~\ref{LSF}-(e) along with the final result~(Fig.~\ref{LSF}-(f)). Then, A polynomial interpolation is used to interpolate the data set into a fixed number of points for all the frames. For the mitral orifice data, as shown in Fig.~\ref{valve} a cubic-spline interpolation coupled with weighted moving average (span=20) for data smoothing was employed.

\begin{figure} 
	\centering
	\includegraphics[width=1.\textwidth]{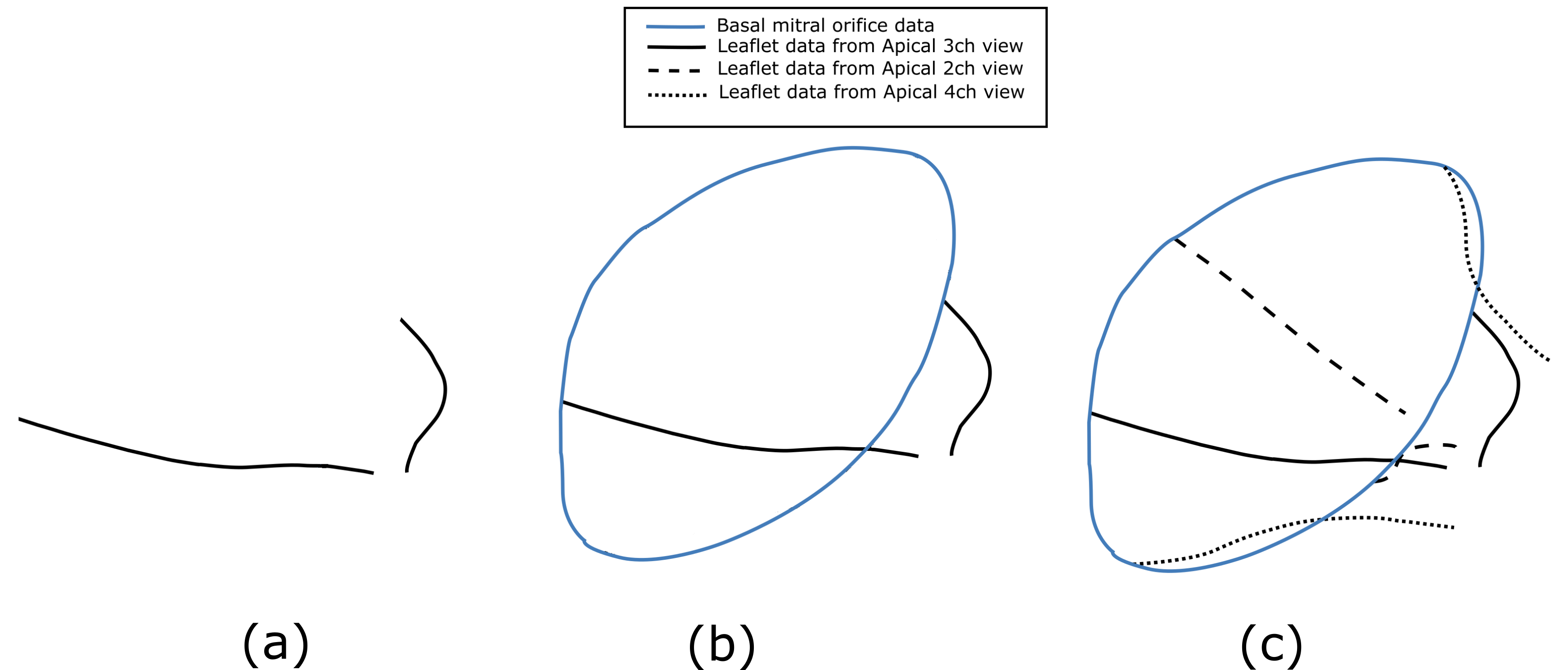}
	\caption{Steps taken for mitral data scaling and orientation. (a) Step-1: Scale the
		mitral leaflet data from the Apical three-chambered view, (b) Step-2: Scale the Basal mitral orifice data based on mitral leaflet data from the Apical three-chambered view, (c) Step-3: Scale the Apical 2ch and 4ch mitral leaflet data.}
	\label{valve_scaling}
\end{figure}

The mitral and aortic orifice data are oriented using angular orientation value of the SA basal from the LV optimization process explained in section~\ref{Scaling}. In addition, The scaling of the mitral data as seen in the 3 LA Apical views~(Fig~\ref{valve_ident}-a to Fig~\ref{valve_ident}-c) and orifice as observed in the SA Basal section~(Fig~\ref{valve_ident}-d) is performed as follows. First the mitral and aortic valve data is scaled using the scaling ratios calculated from the LV reconstruction process~(Fig.~\ref{valve_scaling}-a). Then, the segmented mitral leaflets from LA Apical two-chambered and three-chambered view are fitted to the mitral orifice that is extracted from the SA basal section. For positioning the data, the three-chambered view planar data for the valves as well as the LV is matched to obtain the transverse location of the valve's location. For the longitudinal positioning of the valves, the bases of the valve orifices have been assumed to be $5~mm$ above the LV base section at all time instances~(based on the measurements from the LA sections). For the reconstruction of the aortic valve, only one sectional data is available i.e. from the Apical three-chambered view~(Fig.~\ref{valve_scaling}-a). 

\begin{figure}
	\centering 
	\includegraphics[width=1.0\textwidth]{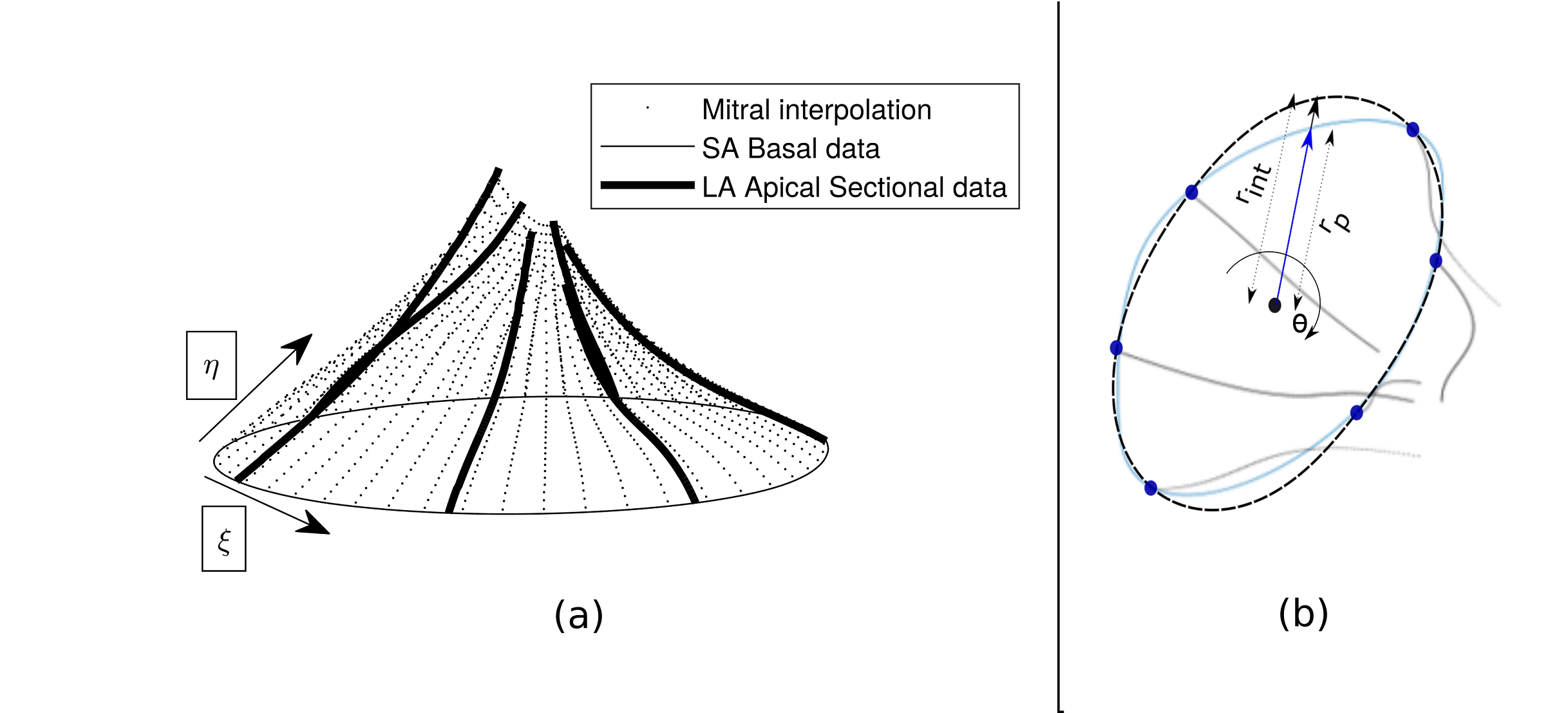}
	\caption{(a) An example showing the 3D spatially interpolated mitral with the sections used for the interpolation scheme, (b) An example showing the prediction and correction shown in Eqn.\ref{pred-corr}}
	\label{MV_rec}
\end{figure}

\begin{figure}
	\centering 
	\includegraphics[width=1.0\textwidth]{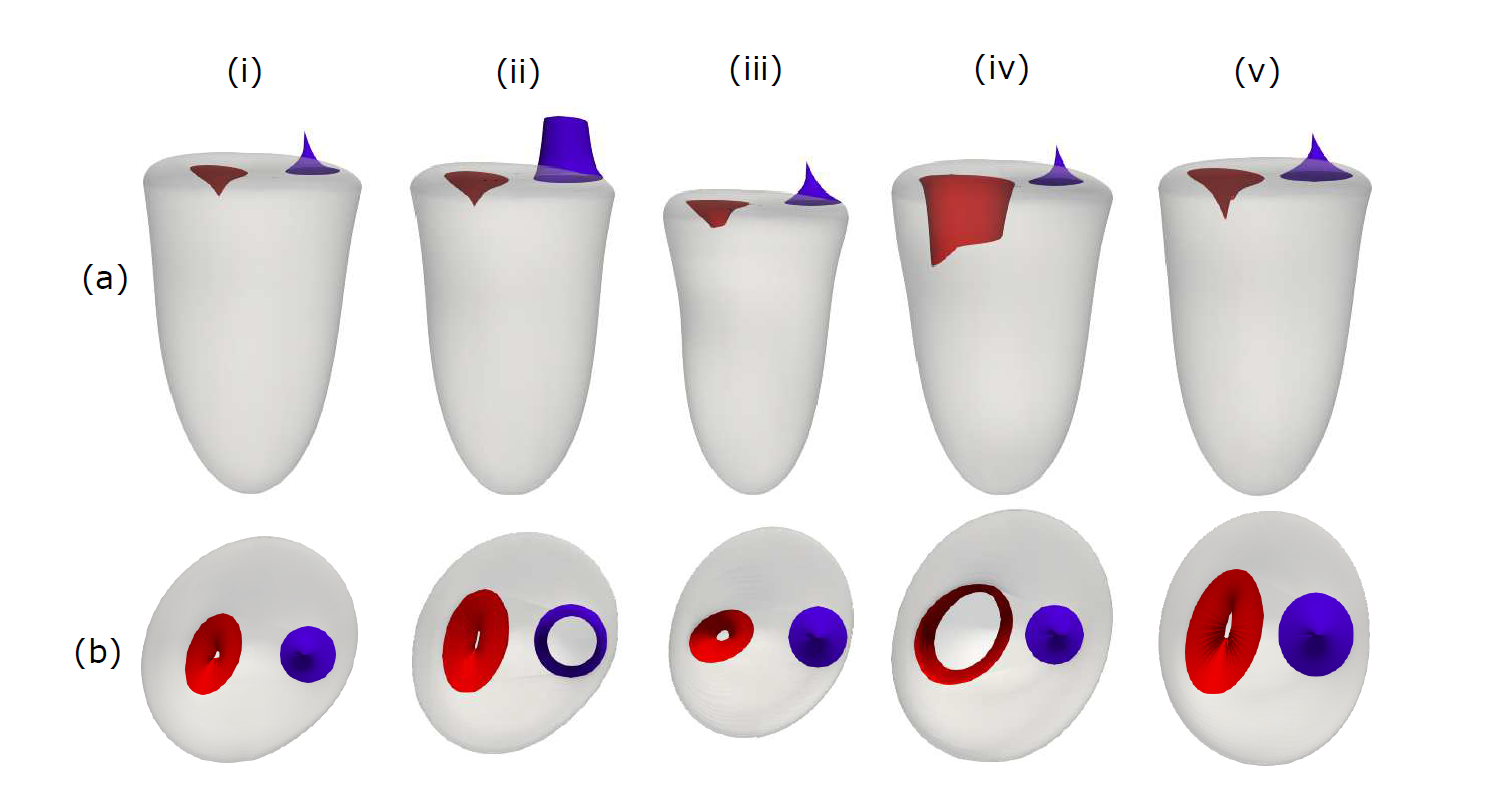}
	\caption{Valve motion along with the left ventricle. (a) Long axis view, (b) Short
		axis view, with the stages given as, (i) early systole (ii) mid systole (iii) end systole (iv) mid diastole (v) end diastole}
	\label{valve_motion}
\end{figure}

In order to reconstruct the mitral valve, a predictor-corrector type of algorithm has been used in this work as follows. The SA sections of the mitral valve are predicted using the LA leaflet data using curve fitting, and after the predictor step, the correction is computed and implemented. Correction at all the LA sections is computed and applied to all the predicted SA sections to have the corrected SA sections for the mitral reconstruction
\begin{equation}
\vec{c}_r(SA_i,LA_j)=~\vec{r}(SA_i,LA_j) \times \frac{\vec{r_p(LA_j)}}{\vec{r_{int}(LA_j)}}
\label{pred-corr}
\end{equation}
where $\vec{c}_r(SA_i,LA_j)$ is the corrected radial dimension, $\vec{r}(SA_i,LA_j)$ is the original dimension, $\vec{r}_p(LA_j)$ is the mitral orifice data, and $\vec{r_{int}(LA_j)}$ is the interpolated data in the mitral orifice section, or the predicted data. The results of this procedure can be found in Fig.~\ref{MV_rec}. Considering two directions, $\xi$ and $\eta$, shown in Fig.~\ref{MV_rec}, the interpolation is performed in one of the directions (e.g. $\xi$), and the correction to the interpolation is done in the other direction (e.g. $\eta$) using a bi-variate interpolation. However, there is only one cross-section~(Apical three-chambered view) of echo images in which AV is clearly visible. Therefore, the AV is reconstructed using only this view assuming that AV has a circular cross-section which is shown to be a reasonable assumption for AV reconstruction~\cite{labrosse2006geometric}. The final reconstruction of the MV and AV during a cardiac cycle is shown in Fig.~\ref{valve_motion} which shows the valve motion is well synchronized with the phases of LV.

\subsection{Superior wall reconstruction} \label{Superior wall reconstruction}
\begin{figure}
	\centering 
	\includegraphics[width=1.0\textwidth]{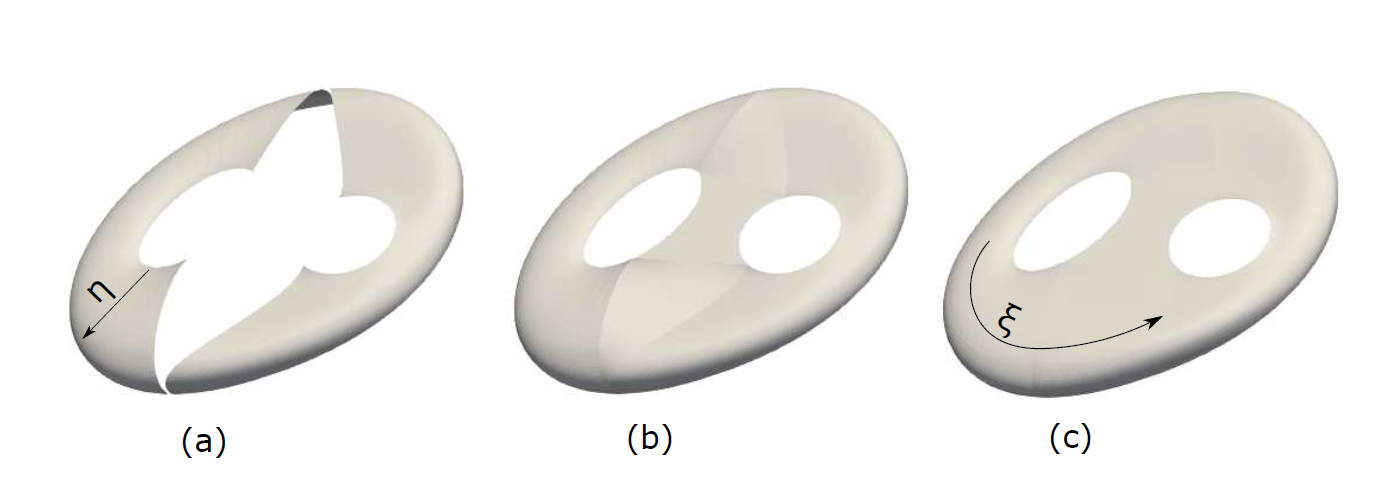}
	\caption{Steps taken for lid generation for a closed geometry. (a) Step-1: Interpolation in $\eta$ direction, (b) Step-2: Linear interpolation to fill spaces, (c) Step-3: Smoothing in $\xi$ direction.}
	\label{lid}
\end{figure}

Closed geometry is essential to perform CFD simulation in the LV. Hence, the superior wall is modeled to connect the valves and the Endocardium wall so as to have a smooth enclosure for the LV geometry. A bivariate strategy is employed for closing the LV geometry with the valves. The bivariate strategy is such that the interpolation is performed in $\eta$ direction, and the smoothing is performed in $\xi$ direction~(Fig.~\ref{lid}) using following steps: 1) a section of the valve bases are connected to the LV base using cubic spline interpolation method 2) the remaining space is covered with linearly interpolated data so as to have a closed geometry 3) smoothing is performed in the orthogonal direction to have a smooth geometry and to incorporate information from both the directions. Steps of this process can be seen in Fig.~\ref{lid}.

\subsection{Hybrid echo-CFD framework} \label{Hybrid echo-CFD framework}
To handle the complex shape and motion of LV, a curvilinear immersed boundary (CURVIB) method~\cite{gilmanov2005hybrid,ge2007numerical,borazjani2008curvilinear} is used which has been extensively validated for a variety of complex flow problems~\cite{borazjani2013parallel,asgharzadeh2017newton} and implemented in various applications such as cardiovascular flows~\cite{Asgharzadeh2019,hedayat2019comparison,hedayat2017platelet}, aquatic motions and vortex dynamics~\cite{asadi2018scaling,daghooghi2016self} and rheology of suspensions~\cite{daghooghi2015influence,daghooghi2018effects}. The 3D reconstructed surfaces of LV from 2D-echo which are meshed using triangular elements are given to the flow solver as an input for each time step to classify the background domain's nodes into fluid, boundary, and solid using an efficient ray tracing algorithm~\cite{borazjani2008curvilinear} which can handle thick, closed-surface bodies. To classify the nodes corresponding to heart valves, which are provided as thin structures, an immersed boundary nodes classification algorithm for thin bodies is used~\cite{borazjani2013fluid}. After immersed boundary nodes classification, the boundary conditions on the solid/fluid interface are reconstructed using the velocity of the 3D reconstructed LV surface using no-slip condition. Finally, the blood flow is driven from/into LV using the volume flux equal to the volumetric change of the LV. The simulations are performed for two cardiac cycles to let the flow reach a quasi-steady state.

\subsection{Computational setup} \label{Computational setup}
\begin{figure}
	\centering 
	\includegraphics[width=1.0\textwidth]{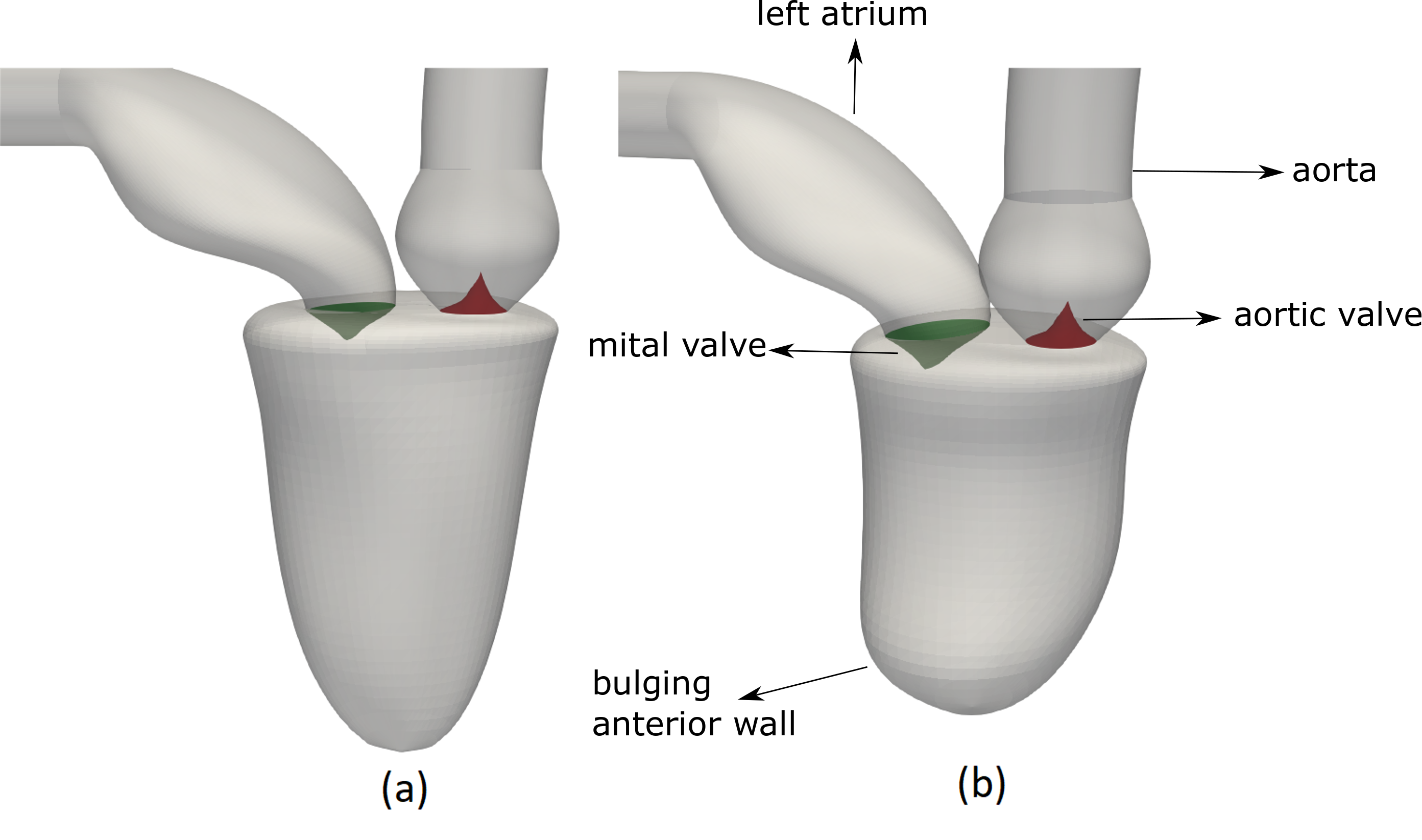}
	\caption{The final reconstructed LV based on echo attached to approximately reconstructed aorta and atrium surfaces for a) healthy and b) AMI cases}
	\label{complete_setup}
\end{figure}

To minimize the influences of the inlet/outlet boundary conditions on the flow inside the LV, simplified surfaces~(not from any medical images) are generated to model left atrium and aorta. These geometries are generated in a way to make the dimensions of the surfaces have realistic values in comparison to the reconstructed LV based on data measurements in previous works \cite{vriz2014normal,otani2016three}. The complete reconstructed geometry used for CFD simulations can be seen in Fig.~\ref{complete_setup}. It is worth mentioning that only one of the aorta and atrium exist in the geometry at the same time~(the aorta just exist during the systole phase and will be removed during diastole and an opposite pattern for left atrium).

The blood flow flux from the left atrium and aorta to/from the LV chamber is specified based on the volumetric change of the LV~(The volumetric flux of LV is calculated based on the change of LV volume in two consecutive time steps). In addition, the velocity boundary condition at the inlet of the atrium is assumed to be uniform. The Navier-Stocks equation is non-dimensionalized with a diameter, $D = 16.38 ~mm$, of the aortic orifice and the bulk velocity, $U = 0.598~m/s$, with a time step $dt = 0.0109~s$ over 2500 time instances during a cardiac cycle. Considering the blood viscosity to be $\nu= 3.3\times 10^{-6} m^2/s$  leads to a Reynold's number of 2950 for the simulations. The LV geometry is discretized with approximately $50,000$ unstructured triangular mesh points and is immersed in a background grid with a dimension of $5.19D \times 3.33D \times 6.32D$~(where $D$ is the aortic diameter) discretized with $161 \times 121 \times 201$ grid points in x, y and z directions, respectively.

\section{Results and discussion}
Here, the effect of fixed apex assumption on the final surface reconstruction and LV flux is studied. In addition, the sensitivity of the results~(reconstruction and CFD simulations) to the smoothing algorithm is investigated. Furthermore, the effect of ventricular valves especially the MV on the flow field in the LV is studied. Finally, the flow hemodynamics of healthy and AMI LVs are compared with each other. The presented results are for healthy LV based on VSWMA smoothing algorithm unless it is mentioned otherwise.


\subsection{Effect of change in positioning and orientation of the LV sections on LV geometry}
\begin{figure} 
	\centering
	\includegraphics[width=0.8\textwidth]{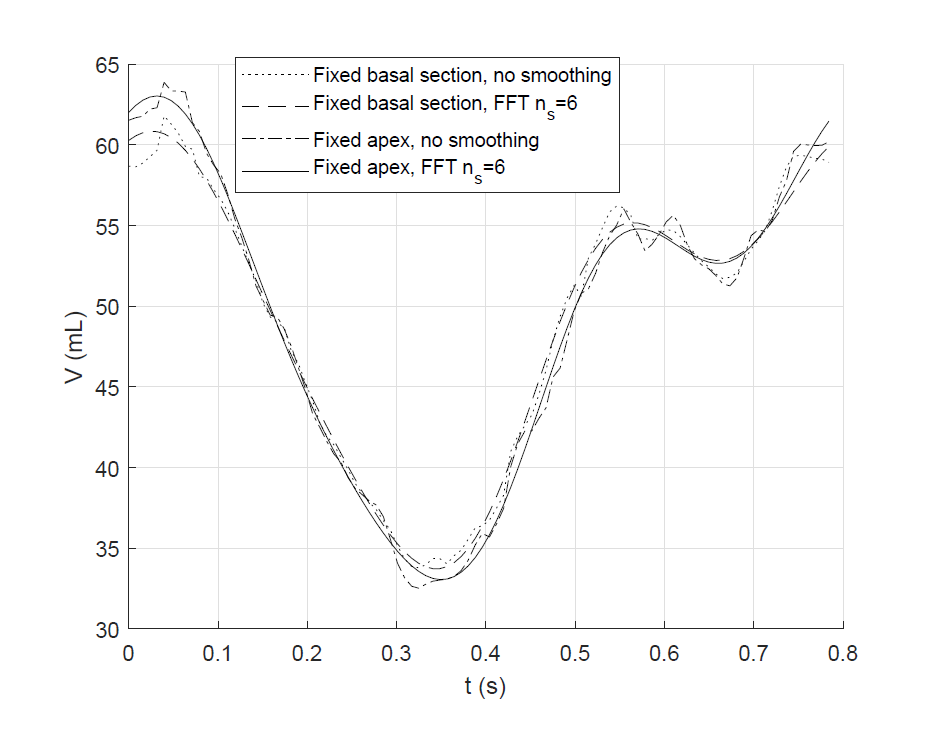}
	\caption{The final reconstructed LV based on echo attached to approximately reconstructed aorta and atrium surfaces for a) healthy and b) AMI cases}
	\label{orientation-effect}
\end{figure}

As previously mentioned, the assumption of the fixed apex is more realistic compared to fixed SA basal section since the apex has the least displacement in LV geometry during a cardiac cycle~\cite{rogers1991quantification}. This assumption will affect the final reconstructed geometry and volumetric flux of LV during a cardiac cycle. Figure~\ref{orientation-effect} shows the volumetric curve of the reconstructed LV with the assumption of having the apex fixed compared to the one with SA basal section fixed at various time instances during a cardiac cycle. The difference in LV flux based on these two assumptions can clearly be seen in this figure. Comparing the ejection fraction~($EF$), which is typically used by physicians as an indicator of heart functionality, an increase of $4\%$~($0.46$ for apex fixed vs. $0.44$ for SA basal section fixed) in the calculation of $EF$ is observed using the apex fixed assumption. The $4\%$ difference in the $EF$ is not be considerable for identifying heart failure with reduced $EF$~(for human usually $EF$ is usually more than $50\%$ and less than $40$ for healthy heart and heart failure, respectively). However, assumption of fixed apex can affect the investigation of flow features inside the LV as well as the LV flux during a cardiac cycle and consequently synchronous of valves opening and closing with the LV flux.

\subsection{Accuracy vs. smoothness in geometric reconstruction} \label{smoothness in geometric}
The segmented data from the echo images are not smooth. Employing different smoothing algorithms can change the results of the final geometric reconstruction of LV. The ideal smoothing algorithm should preserve the shape of data~(segmented from echo) while avoiding over-fitting~(ringing) in the final geometry. Fig.~\ref{smoothing}-a shows the final smoothed curve that resulted from implementing different smoothing algorithms and also different spans for the weighted moving average technique on data extracted from the three-chambered apical LA section. As it can be seen the WMA method with higher weighting span~($20$) results in a length decrease of about $15\%$ in LA section while WMA10 and the b-spline methods result in ringing and instabilities near the SA basal region, which is clearly visible close to the end of systole~(Fig.~\ref{smoothing}-b). However, using variable span moving average with a higher span value near the basal section, unlike the b-spline method, can take care of the instabilities in this region, while using smaller span near the apex provides less smoothing in apex area~(which results in better preservation of the shape of LV). Table~\ref{table_comparison} quantitatively compares the effect of smoothing algorithm on the surface reconstruction process in terms of preserving the important features of the shape of LV~(length of the long axis) and least square error for curve fitting. Our results show the VWMA method provides the best trade-off between the shape preservation and preventing ringing in the final reconstructed geometry.

\subsection{Comparison of LV parameters with their corresponding physiological range}
\begin{figure} 
	\centering
	\includegraphics[width=0.8\textwidth]{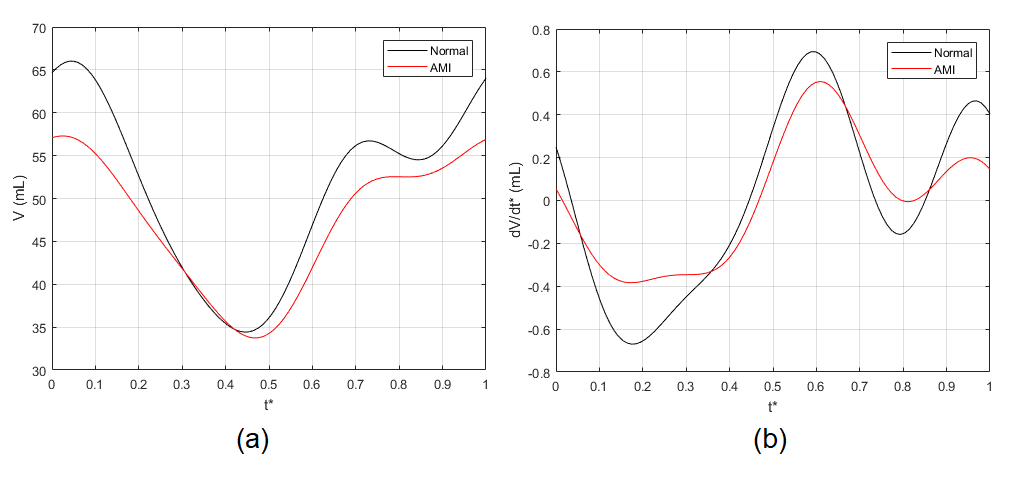}
	\caption{Comparison of a) the volume (V) and b) flux (dV/dt) for Healthy and AMI afflicted pocine LV during a cardiac cycle.}
	\label{healthy-AMI-flux}
\end{figure}

Several parameters related to functionality of LV from reconstructed geometry are compared to their corresponding physiological ranges to show that the reconstruction from echo images is comparable to physiological data. The volume and volumetric flux versus time for both healthy and AMI left ventricles are presented in Fig.~\ref{healthy-AMI-flux}. The time is non-dimensionalized in this figure so as to have the same time duration for both the healthy and AMI cases. Various parameters including $EF$, the ratio of maximum 
fluxes during the E-wave and A-wave~($E/A$ Ratio), Deceleration Time~($DT$), Stroke volume~($SV$) and Cardiac Output~($CO$) are calculated based on the curves in Fig.~\ref{healthy-AMI-flux}. Table~\ref{table-parameters} shows the comparison between the calculated parameters and their physiological ranges reported in previous in vivo experiments for a porcine LV~\cite{reiter2016early}. As it can be seen in this table, the parameters calculated here for healthy LV lie within the physiological ranges of in-vivo experiments. However, the $E/A$ ratio, $SV$ and $CO$ for the AMI afflicted subject's LV lie outside the physiological ranges for normal subject's LV as can be expected.

\subsection{CFD simulation for the reconstructed LV assembly}
\begin{figure} 
	\centering
	\includegraphics[width=0.5\textwidth]{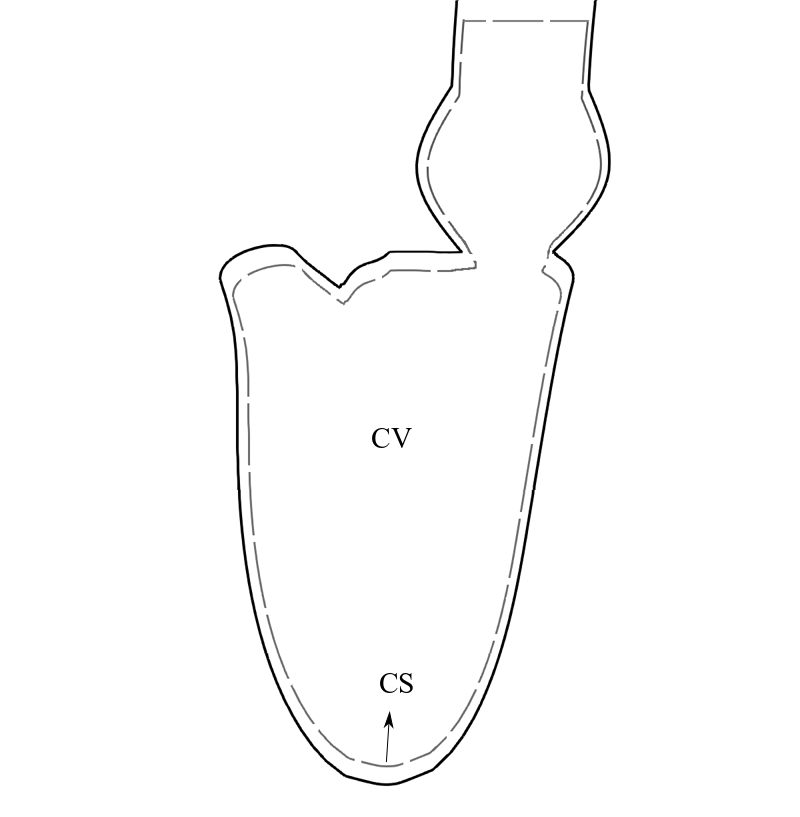}
	\caption{The schematic control volume and control surface used for calculating energy loss in the left ventricle on the mid-plane passing through the aorta}
	\label{control}
\end{figure}

It is well known that the blood flow pattern in the LV has a direct impact on the heart performance~\cite{seo2013effect}. However, this flow pattern is directly related to the accuracy of LV reconstruction and valves' motion. Hence, in this section, the impact of ventricular valves and different smoothing algorithm for LV reconstruction as well as LV dysfunction~(AMI) on the performance of LV in terms of energy loss during the whole cardiac cycle is measured using the energy equation for a control volume as follows:

\begin{equation}
\frac{dE}{dt}=\dot{Q}-\dot{W}=\frac{\partial}{\partial t}\int_{CV}{\rho e~dV}+\int_{CS}{(\rho e+p)\vec{V}.\vec{dA}}
\end{equation}
where $E$ is the total energy of the system~(Fig.~\ref{control}), $\dot{Q}$ is the rate of heat transfer to the system, $\dot{W}$ is the rate of work done on/by the system, $CV$ is the control volume, $CS$ is control surface, $\vec{V}$ velocity of flow, $p$ is pressure, $\rho$ is blood density, and $e$ is the energy per unit mass 
\begin{equation}
e=u+V^2/2+\vec{g}z
\end{equation}
where $u$ is the internal energy of the fluid, $V^2/2$ is the kinetic energy. Integrating over a cardiac cycle assuming that the system has reached the quasi-steady state and neglecting the heat transferred to the LV as well as gravity and difference of internal energy of inlet and outlet, the above equation reduces to

\begin{equation}
loss= \int_{0}^{T}{\int_{CS}{\rho~(p/{\rho}+V^2/2)~\vec{V}. \vec{dA}}}
\end{equation}
where $T$ is time at the end of the cardiac cycle. Since AMI and Healthy left ventricle have different stroke-volume and heart beat rate, to be able to compare the performance of the LV in different cases the rate of loss is calculated per $lit$ of blood pumped in each simulation as follows:
\begin{equation}
\dot{loss}~ (J/lit)= \frac{loss \times HR}{CO}
\end{equation}
where $CO$ is the cardiac output and $HR$ is the heart rate. Table~\ref{work_load} compares the $\dot{loss}$ of the LV during a cardiac cycle for different smoothing algorithms, without heart valve, and AMI simulations.  

\subsubsection{Effect of mitral valve}
\begin{figure} 
	\centering
	\includegraphics[width=1.0\textwidth]{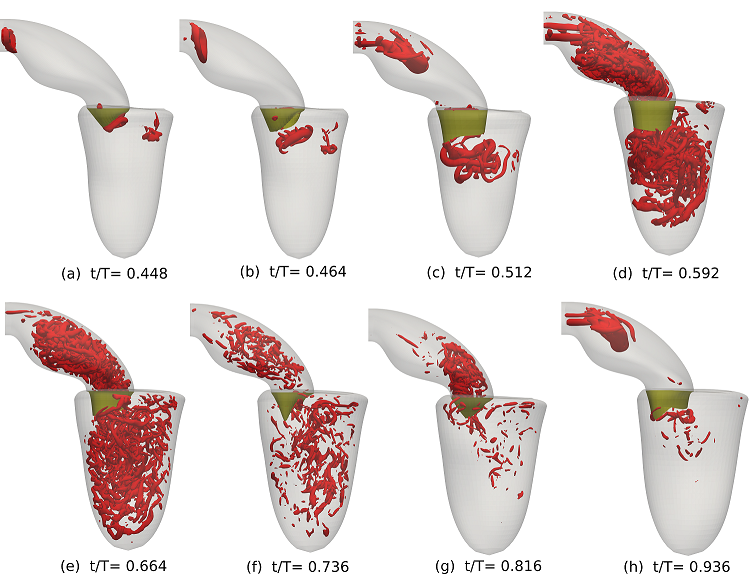}
	\caption{3D vortical structures visualized at different time instances during the
		diastolic filling using the iso-surfaces of q-criteria for a healthy subject's LV with re-constructed valves for different time instants during diastole~(t/T=instance time / cardiac cycle length)}
	\label{healthy-vortex}
\end{figure}

\begin{figure} 
	\centering
	\includegraphics[width=1.0\textwidth]{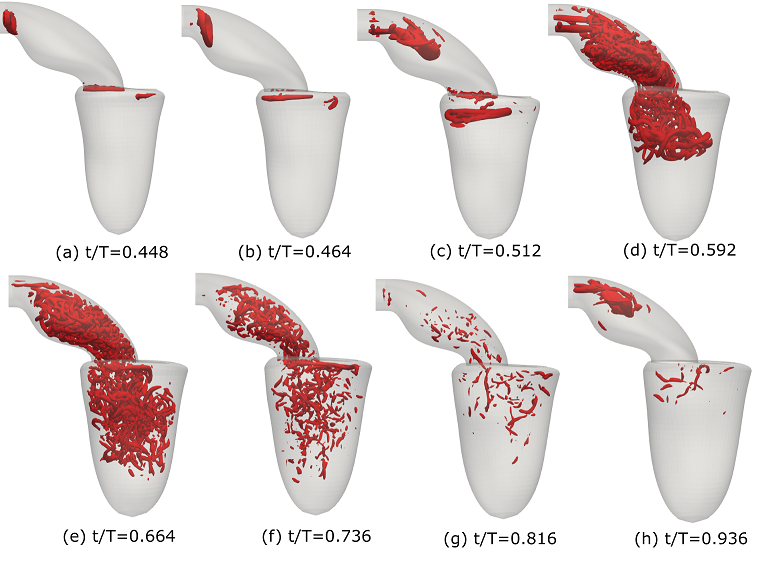}
	\caption{3D vortical structures visualized at different time instances during the
		diastolic filling using the iso-surfaces of q-criteria for a healthy subject's LV without valves for different time instants  during diastole~(t/T=instance time / cardiac cycle length)}
	\label{healthy-vortex-without}
\end{figure}

The vortex ring generated during the rapid filling~(E-wave) is one of the key characteristics of intraventricular flows~\cite{gharib2006optimal}. Figure~\ref{healthy-vortex} shows the vortical structures inside healthy LV with incorporate ventricular valves during diastole phase visualized by iso-surface of q-criteria \cite{haller2005objective}. For comparison, Fig.~\ref{healthy-vortex-without} shows the same visualization for the LV without valves at the same time instances. As can be observed in Fig.~\ref{healthy-vortex} at time $t/T=0.448$~(t/T=instance time / cardiac cycle length), in the early diastolic phase where the mitral valve leaflets are just beginning to open, the vortex ring starts forming on the tip of the leaflet of the mitral valve. Since the mitral orifice is a circular this vortex ring has a circular shape. The vortex ring starts to pinch off and propagates inside the LV around the peak E-wave~(Fig.~\ref{healthy-vortex}-b). Due to the asymmetric geometry of mitral leaflets, the ring propagates towards the posterior wall of LV while starting to disintegrate as it approaches the wall. This ring finally hits the wall and begins to break down into small-scale vortical structures which fill the whole volume of LV. During the A-wave also another vortex ring is generated. However, this time the ring is weaker and dissipates faster without propagating much in the LV. 

Comparing the q-criteria visualization of LV without the valve; it can be clearly observed that the vortex ring starts forming at the mitral annulus~(Fig.~\ref{healthy-vortex-without}). Due to the absence of the mitral valves, the symmetric ring propagates towards the apex of LV. However, since the mitral annulus has a larger orifice area compared to the orifice of the MV, the ring is weaker and the core of the vortex has smaller propagation speed. The peak velocity near the mitral annulus in LV with MV is around $1.37~m/s$ which is in agreement with the previously published physiological values for a healthy LV~\cite{sotiropoulos2016fluid}, whereas in the simulation without MV this value is $0.76~m/s$. Therefore, the vortex ring is not traveling far in the apical direction inside the LV and it starts breaking down after propagating about $30\%$ of the LV length. This shallow vortex ring penetration depth that happens in the simulations without MV, can negatively affect the washout ability of LV in the apex region. In addition, as previously shown by Seo et al.~\cite{seo2014effect} the presence of MV results in a higher asymmetric diastolic flow pattern and consequently a counter-clockwise~(CCW) circulation which increases the washout potential of LV. This is in agreement with our results which show the CCW circulation of about $63~ cm^2/s$ and $41 ~cm^2/s$ for LV with and without MV, respectively, which is an increase of $35\%$. As it been shown in the previous study by Seo et al.~\cite{seo2013effect}, the asymmetric flow pattern during the diastole phase which is also reflected in the higher CCW circulation will increase the efficiency of blood ejects towards aorta during the systole phase. In addition, effect of AV and MV on the $\dot{loss}$ is investigated in Table~\ref{work_load}. As it can be seen the presence of ventricular valve's AV increases the $\dot{loss}$. This increase is mainly due to the presence of AV during the systole. However, the presence of AV is essential to prevent backflow during the diastolic phase.   

\subsubsection{Sensitivity of the numerical simulations to smoothing algorithms} \label{numerical simulations smoothing}
The qualitative comparison of ring vortex formation and its propagation using q-criteria shows no significant difference in the flow pattern inside the LV for different smoothing algorithms. In order to make a more quantitative comparison, the $\dot{loss}$ for different smoothing is calculated. Again, results do not show a significant variation in $\dot{loss}$ due to different smoothing algorithms. Thus it can be concluded the although the choice of smoothing algorithm can have some impact on the accuracy of the reconstruction and thus local flow pattern~(in the regions near the ringing in the final geometry), the main flow patterns are not very sensitive to the smoothing method. 

\subsubsection{Comparing healthy and AMI reconstructed LV} \label{Comparing healthy and AMI}
\begin{figure} 
	\centering
	\includegraphics[width=1.0\textwidth]{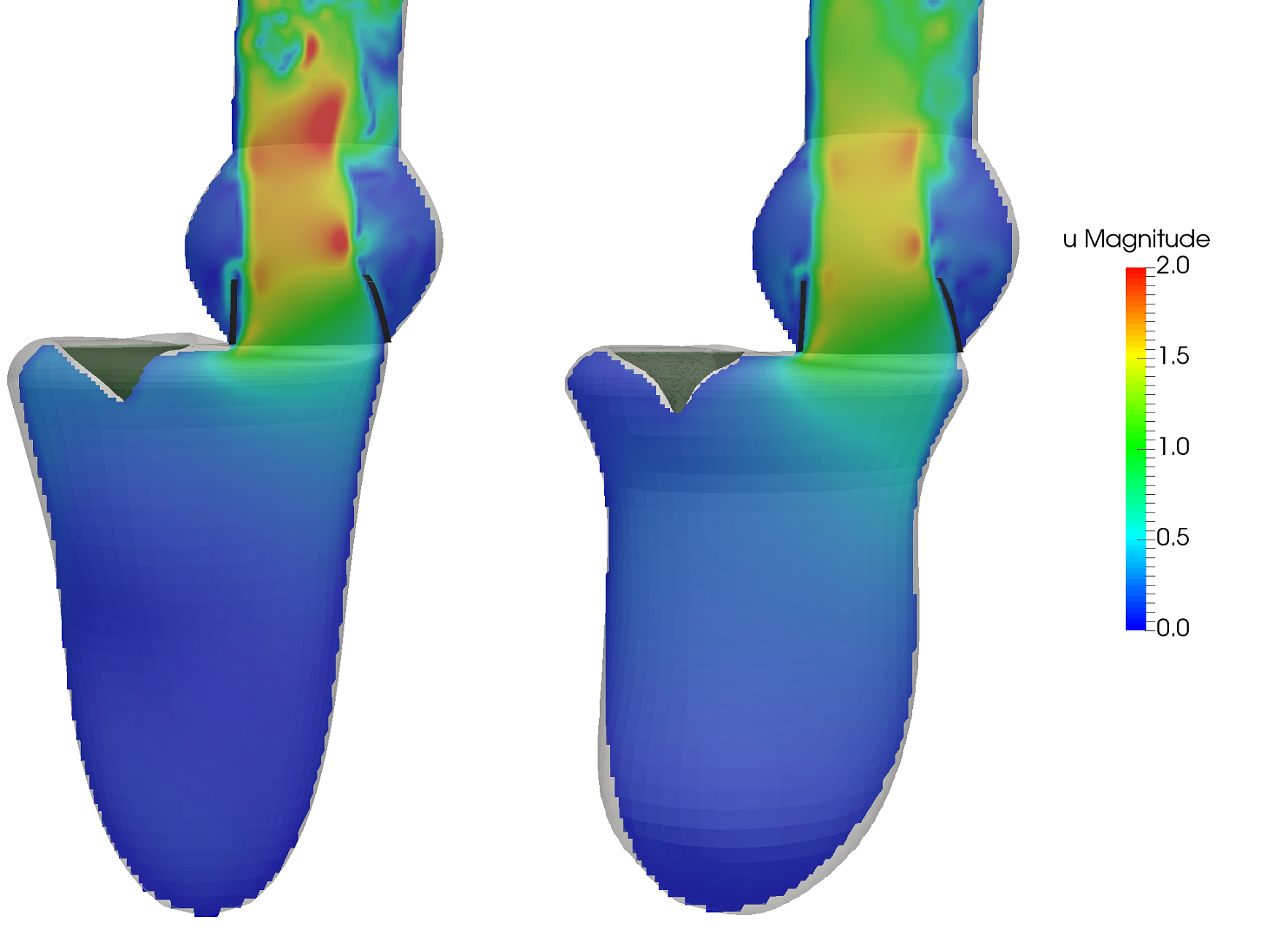}
	\caption{Comparison of the flow velocity in the aortic orifice at peak systole between (a) Healthy (b) AMI left ventricle }
	\label{AMI-healthy_velocity}
\end{figure}

AMI can significantly affect the performance of LV during the systole. Several studies~\cite{white1987left,st1994quantitative,moller2003prognostic} investigated the effect of the AMI on systolic functionalities in terms of $EF$ and the amount of blood flux through the aorta. Figure~\ref{AMI-healthy_velocity} shows the maximum velocity of flow through the aortic orifice at peak systole. As it can be seen, the magnitude of velocity in the healthy LV is higher than the AMI case which is due to lower $CO$ and $EF$ in the LV with AMI compared to the healthy one. 

\begin{figure}
	\centering 
	\includegraphics[width=1.0\textwidth]{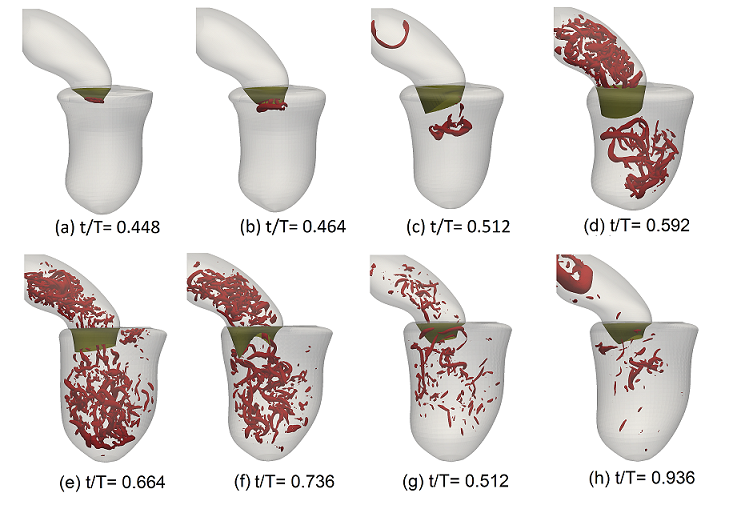}
	\caption{3D vortical structures visualised at different time instances during the
		diastolic filling using the iso-surfaces of q-criteria for a AMI afflicted subject's LV~(t/T=instance time / cardiac cycle length) }
	\label{AMI-vortex}
\end{figure}

AMI can also affect the diastolic performance of the LV~\cite{moller2006prognostic} in terms of the filling pattern. A bulge or dyskinetic region in the reconstructed LV surface of the AMI-afflicted subject can be clearly seen in Fig.~\ref{AMI-vortex}. The visualization of vortical structures using q-criteria in this figure shows the formation of a vortex ring from the tip of mitral leaflets the same as the healthy subject. However, there is a time delay for vortex formation/propagation in the AMI versus the healthy LV which is due to delay in the starting time of diastolic phase in AMI LV that can be clearly seen in Fig.~\ref{healthy-AMI-flux}. In addition, the vortex ring formed in the AMI subject is weaker and has a lower propagation velocity($1.12~m/s$ compared to the healthy one $1.37~m/s$) and thus it disintegrates and dissipates in the LV sooner. It can also be seen that in the AMI simulation the vortical structures are predominantly found in the region directly beneath the mitral annulus, as compared to a more uniform and looped sweeping of structures in a healthy subject's LV. Comparison of the $\dot{loss}$ in Table~\ref{work_load} for AMI and healthy LV shows that the AMI left ventricle has higher $\dot{loss}$ by approximately $20\%$ compared to all healthy LVs with different smoothing. Our results suggest that $\dot{loss}$ can be used as a promising indicator to measure the performance of LV.


\section{Conclusions}
An improved methodology based on the previous work of Rajan et al.~\cite{rajan2016automated} for the automated 3D reconstruction of the LV from multiple axis echo of both long-axis and short-axis sections has been implemented. To the best of our knowledge, this is a first attempt made to perform reconstruction of a healthy and an AMI LV along with mitral and aortic valves using porcine 2D echo~(potentially usable for patient-specific reconstructions) obtained using standard cross-sectional views. The sensitivity of the CFD simulation to the process of reconstruction was investigated. The results show that the choice of smoothing algorithm for image segmentation can slightly affect the final reconstructed geometry in terms of generating non-physical ringing in the wall of the; however, the comparison of energy loss and flow visualizations in CFD simulations with different smoothing algorithms shows the choice of smoothing does not change the main flow pattern during the cardiac cycle. Furthermore, the change in positioning and orientation of the LV sections has also brought the reconstruction one step closer to a more physical reconstruction with the LV apex fixed in space along with the short-axis sections. Having reconstructed the LV from the multiple axes 2D echo data, the same data was used to extract mitral and aortic valves. Our results also suggest the absence of MV and AV which is the case in most of the previous LV simulations can change the flow pattern inside the LV in a non-physiological way in terms of vortex formation/propagation, flow circulation during the diastole phase and energy loss. Our results show the vortex ring formed in the healthy is stronger and has faster propagation speed compared to AMI which may suggest a better apex washout for a healthy subject. In addition, our results suggest the energy $\dot{loss}$ can be a good indicator for the performance of the LV in the numerical simulations.

\section*{CONFLICT OF INTEREST STATEMENT}
The authors declare that there is no conflict of interest regarding the publication of this article.


\section*{ACKNOWLEDGEMENT}
This work was supported by American Heart Association grant 13SDG17220022 and the Texas A$\&$M High Performance Research Computing center~(HPRC).

\bibliography{wileyNJD-AMA}%

\begin{table}[t]
	\caption{Comparison of smoothing algorithms}
	\begin{center}
		\label{table_comparison}
		\begin{tabular}{c l l}
			\hline
			Algorithm & Length change~(\%) & $L2$ error \\
			\hline
			b-spline & 3.1 & 66.78  \\
			WMA10 & 0.3 & 12.11  \\
			WMA20 & 8.22 & 108.9  \\
			VSWMA & 3.78 & 35.79  \\
			\hline
		\end{tabular}
	\end{center}
\end{table}

\begin{table}[t]
	\caption{Comparison of LV parameters from the reconstructed LV~(healthy and AMI)
		compared to the physiological ranges}
	\begin{center}
		\label{table-parameters}
		\begin{tabular}{c l l l}
			\hline
			& Physiological ranges (Porcine LV) & Healthy LV & AMI LV\\
			\hline
			HR (bpm) & 65-100 & 80 & 77\\
			EF & $0.52 \pm 0.03$ & 0.47 & 0.43\\
			E/A~Ratio & $1.11 \pm 0.28$ & 1.38 & 2.5 \\
			DT~(/cycle) & $0.2 \pm 0.02$ & 0.236 & 0.23\\
			SV~(mL) & $29 \pm 5$ & 29 & 24\\
			CO~(L/min) & $2.35 \pm 0.21$ & 2.3 & 1.85\\
			\hline
		\end{tabular}
	\end{center}
\end{table}

\begin{table}[t]
\caption{Comparison of LV $\dot{loss}$ for different test cases}
\begin{center}
\label{work_load}
\begin{tabular}{c l l}
\hline
case (smoothing) & $\dot{loss}~(J/lit)$ \\
\hline
Healthy (b-spline) & 8.44   \\
Healthy (WMA10) &  8.51  \\
Healthy (WMA20) & 8.32  \\
Healthy (VSWMA) & 8.76  \\
Healthy without valve (VSWMA) & 3.89  \\
AMI (VSWMA) & 10.3   \\
\hline
\end{tabular}
\end{center}
\end{table}

\end{document}